\documentclass[elsart12,epsf,eqsecnum,graphics,cite,nofootinbib]{revtex4}

\usepackage{epsfig,epsf}
\usepackage{graphicx,color}
\usepackage{grffile}
\usepackage{amsmath,amsfonts,amssymb,amsthm,nccmath,latexsym,mathtools} \usepackage[mathscr]{euscript}

\def\p{{\mathbf p}}

\def\q{{\mathbf q}}

\def\k{{\mathbf k}}
\def\x{{\mathbf x}}

\def\K{{\mathbf K}}

\def\P{{\cal P}}

\newcommand{\beq}{\begin{eqnarray}}
\newcommand{\eeq}{\end{eqnarray}}
\newcommand{\be}{\begin{eqnarray*}}
\newcommand{\ee}{\end{eqnarray*}}
\DeclareMathOperator{\tr}{tr}

\begin{document}


\voffset1.5cm

\title{Hanbury-Brown-Twiss measurements at large rapidity separations, or can we measure the proton radius in p-A collisions?}

\author{ Tolga Altinoluk$^1$, Nestor Armesto$^1$, Guillaume Beuf$^2$, Alex Kovner$^3$ and Michael Lublinsky$^{2}$}

\affiliation{
$^1$ Departamento de F\'{i}õsica de Part\'{i}culas and IGFAE, Universidade de Santiago de Compostela, 15706 Santiago de Compostela, Galicia-Spain\\
$^2$ Department of Physics, Ben-Gurion University of the Negev,
Beer Sheva 84105, Israel\\
$^3$ Physics Department, University of Connecticut, 2152 Hillside
Road, Storrs, CT 06269-3046, USA}

\date{\today}

\begin{abstract}
We point out that current calculations of inclusive two-particle correlations in p-A collisions based on the Color Glass Condensate approach exhibit a contribution from Hanbury-Brown-Twiss correlations. These HBT correlations are quite distinct from the standard ones, in that they are apparent for particles widely separated in rapidity. The transverse size of the emitter which is reflected in these correlations is the gluonic size of the proton. This raises an interesting possibility of measuring the proton size directly by the HBT effect of particle pairs produced in p-A collisions.
\end{abstract}

\maketitle
\section{Introduction}
The observation by the CMS \cite{CMS}  and ATLAS \cite{Aad:2015gqa} collaborations of ridge correlations in  high multiplicity p-p collisions at the Large Hadron Collider (LHC) has triggered a wave of theoretical attempts to understand the nature of such correlations. Similar correlations have been subsequently observed by all three large LHC experiments in p-Pb collisions \cite{CMS:2012qk}, and much more detailed studies of the properties of these correlations are available today. 

There are indications that the origin of these correlations is due to collective (hydrodynamic?) behaviour of the system produced in the collision \cite{Bozek:2012gr}. However, a good quantitative description of the data is also achieved \cite{DV} within the Color Glass Condensate (CGC) based ``glasma graph'' approach \cite{DMV,ddgjlv}, which ascribes the origin of the correlations entirely to the structure of the initial state. In reality it is likely that both mechanisms are contributing to the correlations, probably in different transverse momentum ranges. At any rate, it is interesting to have a deeper understanding of the initial state induced correlations. Note that other explanations within the CGC \cite{correlations,LR} and in other frameworks  \cite{Hwa:2008um} also exist, but they will not be touched upon in this paper.

In a recent paper \cite{BE} we have shown that the bulk of the correlations in the glasma graph approach originates from the Bose enhancement of gluons in the incoming projectile and target wave functions. This quintessential quantum mechanical effect increases the probability to find gluons with the same transverse momentum in the wave function of the incoming projectile (and target). The scattering smears the momentum of the gluons to some degree, but significant correlations in the gluons produced in the final state are nevertheless observable in favourable kinematics. In this note we point out to an additional physical effect present in the glasma graph calculation: Hanbury-Brown-Twiss (HBT) correlations between gluons far separated in rapidity (for reviews on interferometry studies in heavy-ion collisions see \cite{Weiner:1999th}). This leads to a potentially observable effect in the final state mesons which may allow a direct measurement of gluonic ``size'' of the proton. The HBT signal is also interesting in that it correlates gluons with same and opposite transverse momenta (in analogy with the double ridge structure of  \cite{DV}).

The fact that the CGC approach contains HBT correlations is not new and has been recognised before \cite{adrian,yury}. The aim of this note is to understand their unique features, in particular their long range in rapidity nature and  the fact that they reflect the gluonic size of the proton, within the standard setup in HBT studies \cite{Heinz:1996bs}. We also make contact with the glasma graph approach.

This paper is organised as follows. In Sec. \ref{basics} we present a short review of HBT correlations. Here we essentially follow the excellent reviews \cite{Heinz:1996bs} with some slight change of notation. In Sec. \ref{cgc} we discuss HBT correlations in a system of gluons emitted from a Lorentz contracted source, and point out that in this case the HBT correlations extend over a long range in rapidity. We also discuss the typical transverse structure of the emitter expected in the CGC approach, and note that in p-A collisions its transverse size is that of the proton $R^2$, while the transverse area of the region from which the emission is coherent is the inverse saturation momentum of the target $Q_s^{-2}$. Thus, if $R^2Q_s^2\gg 1$ the emission is dominated by the HBT signal that measures the proton size. In Sec. \ref{graphs} we translate this qualitative discussion into the language of glasma graphs and show which glasma graphs correspond to the HBT signal discussed above. In Sec. \ref{disc} we offer some concluding comments.

\section{Basics of HBT}
\label{basics}

In this section we review the basics of HBT correlations, following closely the reviews  \cite{Heinz:1996bs}. The object of study is the normalized two-particle correlation function. 
The single and two-particle spectra are defined as 
\beq
\P_1(\p)&=&\, \frac{dN}{d^3p}=\, \langle \hat{a}^{\dagger}_\p\hat{a}_\p\rangle, \\
\P_2(\p_a,\p_b)&=& \, \frac{dN}{d^3p_ad^3p_b}=\, \langle  \hat{a}^{\dagger}_{\p_a} \hat{a}^{\dagger}_{\p_b} \hat{a}_{\p_a} \hat{a}_{\p_b} \rangle,
\eeq
where $ \hat{a}^{\dagger}$ and $ \hat{a}$ are the creation and annihilation operators of particles (usually charged mesons)  with momenta $p_i=(p_i^0,\p_i)$ and $\langle \cdots \rangle $ stands for the averaging over the ensemble of events. 

The two-particle correlation function $C(\p_a,\p_b)$ is defined as 
\beq
C(\p_a,\p_b)=\frac{\langle N \rangle^2}{\langle N(N-1)\rangle} \frac{\P_2(\p_a,\p_b)}{ \P_1(\p_a) \P_1(\p_b)}\ ,
\eeq
where $\langle N\rangle$ is the average number of particles and $\langle N(N-1)\rangle $ the  average number of pairs, such that
\beq
\int d^3p\, P_1(\p) &=& \langle N\rangle, \\
\int d^3p_ad^3p_b\, P_2(\p_a,\p_b)&=& \langle N(N-1)\rangle \, .
\eeq
The averaging over the event ensemble can be represented in terms of a density matrix $\hat \rho$, so that for an arbitrary observable 
$\hat{O}$, one has
\beq
\langle \hat{O} \rangle = \tr ( \hat{\rho} \,\hat{O}).
\eeq

The particles are emitted from the interaction region, and stop interacting once they leave the ``freeze-out surface''.  Once outside the interaction region, the particles propagate freely until reaching the detector.  The standard description of the ``emitter'' is in terms of a classical source $J(x)$ which emits the pions, so that for a given source $J(x)$ the state of the pion field (between the emitter and the detector) is given by
\beq\label{coherent}
| J \rangle= e^{\bar{n}/2}\exp\left(i \int d^3p\, \tilde{J}(\p)\,  \hat{a}^{\dagger}_{\p} \right) | 0\rangle,
\eeq 
where $\tilde{J}(\p)$ is the on-shell Fourier transform of the classical source $J(x)$,
\beq
\tilde{J}(\p)= \int \frac{d^4x}{\sqrt{(2\pi)^32E_{\p}}} \,\exp \left[ i\left( E_{\p}t -\p\cdot\x\right)\right] J(x),  
\eeq
with $E_{\p}=\sqrt{\p^2+m_\pi^2}$ and the normalization of the state $\bar{n}=\int d^3p\, | \tilde{J}(\p)|^2$. This coherent state is an eigenstate of the annihilation operator, 
\beq
\hat{a}_\p |J\rangle = i\tilde{J}(\p)|J\rangle \, .
\eeq

In the case of the emission of the particle from a single coherent source $J(x)$, the density operator of the ensemble is just a projection operator on the coherent state:
\beq\label{dm}
\hat{\rho}=|J\rangle\langle J|\, . 
\eeq
One then obtains
\beq
\P_1(\p)&=&\langle J|\hat{a}^{\dagger}_{\p}\hat{a}_{\p}|J\rangle=|\tilde{J}(\p)|^2,\\
\P_2(\p_a,\p_b)&=& \langle J| \hat{a}^{\dagger}_{\p_a}\hat{a}^{\dagger}_{\p_b}\hat{a}_{\p_a}\hat{a}_{\p_b}|J\rangle= |\tilde{J}(\p_a)|^2|\tilde{J}(\p_b)|^2.
\eeq
The two-particle spectrum is given by the square of the one particle spectrum  and thus there is no Bose-Einstein correlations from the emission off a single coherent source.

A fixed source $J(x)$ corresponds to a single event. Averaging over ensemble of events corresponds to averaging over the ensemble of sources $J$.
In reality different regions of the freeze-out surface are not coherent. They vary from event to event independently of each other.  This situation is modelled by considering the emitter
to be a chaotic superposition of classical sources, whose phases vary in a random uncorrelated way. As a model for such source we take
\beq
J(x)=\sum_{i=1}^N e^{i\phi_i}e^{-ip_i\cdot(x-x_i)}J_0(x-x_i).
\label{ch_J}
\eeq
Here each individual source has been shifted from its original (four-)position by $x_i$ and boosted by momentum $p_i$. This of course corresponds to the individual sources to be spatially separated as well as moving independently of each other. Additionally, a random phase $\phi_i$ is associated with each individual source $J_0(x-x_i)$. The total emitter  $J(x)$ is the sum of all the individual sources. The on-shell Fourier transform of it is written
\beq
\tilde{J}(\p)=\sum_{i=1}^N e^{i\phi_i} e^{ip\cdot x_i} \tilde{J}_0(p-p_i),
\eeq
where 
\beq
 \tilde{J}_0(p-p_i)= \int \frac{d^4x}{\sqrt{(2\pi)^32E_p}} e^{i(p-p_i)\cdot x} J_0(x) \, .
\eeq
The state of the produced particles now depends on the source positions, source momenta and  phases. The ensemble of sources, corresponding to the distribution of events, can be defined in terms of a density operator $\hat{\rho}$ which specifies the distribution of the source parameters. Assuming that the number of sources is distributed with a probability $P_N$, the phases are distributed randomly between $0$ and $2\pi$, and the source positions $x_i$ and momenta $p_i$ are distributed with a density $n(x,p)$ with the following normalization:
\beq
\sum_{N=0}^{\infty}P_N=1,  \ \ \sum_{N=0}^{\infty}NP_N= N_S,\ \ 
\int d^4x\, d^4p\,  n(x,p)=1,
\eeq
the corresponding ensemble average of an arbitrary operator $\hat{O}$ is given by
\beq
\tr( \hat{\rho}\, \hat{O} )= \sum_{N=0}^{\infty} P_N \int  \int_{\phi=0}^{2\pi} D_N[J]\, \langle J [N; \{x,p,\phi\}]|\hat{O}| J [N; \{x,p,\phi\}]\rangle \, ,
\label{avr_O}
\eeq
where the measure $D_N[J]$ is defined as
\beq
D_N[J]=\prod_{i=1}^N d^4x_i \, d^4p_i  \,n(x_i,p_i) \frac{d\phi_i}{2\pi}\,.
\eeq 
Using Eq. \eqref{avr_O}, it is straightforward to calculate the single particle spectrum:
\beq
\langle \hat{a}^{\dagger}_{\p}\hat{a}_{\p}\rangle = \sum_{N=0}^{\infty}P_N  \int  \int_{\phi=0}^{2\pi} D_N[J]\, 
\sum_{n,n'=1}^Ne^{i(\phi_n-\phi_{n'})} e^{ip\cdot(x_n-x_{n'})}\tilde{J}^*_0(p-p_{n'})\tilde{J}_0(p-p_n).
\eeq
Due to the averaging over the phases only the terms $n=n'$ give non-vanishing contribution. Hence, the single particle spectrum reads
\beq
\P_1(\p) =  N_S\int d^4x' \, d^4 p' \,  n(x',p')\,  |\tilde{J}_0(p-p')|^2\equiv  \langle |\tilde{J}(\p)|^2\rangle.
\eeq
Before calculating the two-particle spectrum let us consider
\begin{eqnarray}
\langle \hat{a}^{\dagger}_{\p_a}\hat{a}_{\p_b}\rangle &=& \sum_{N=0}^{\infty}P_N  \int  \int_{\phi=0}^{2\pi} D_N[J]\, 
\sum_{n,n'=1}^{N} e^{i(\phi_n-\phi_{n'})} e^{i(p_b\cdot x_n-p_a\cdot x_{n'})}\tilde{J}^*_0(p_a-p_{n'})\tilde{J}_0(p_b-p_n)\nonumber\\
&=&N_S \int d^4x' \, d^4p' \, n(x',p') e^{i(p_b-p_a)\cdot x'}\tilde{J}^*_0(p_a-p')\tilde{J}_0(p_b-p')\,\equiv\, \langle J^{*\, a}(\p_a) J^b(\p_b)\rangle.
\label{adaga}
\end{eqnarray}
Using Eq. \eqref{adaga}, we get for the two-particle spectrum
\beq\label{4a}
\langle \hat{a}^{\dagger}_{\p_a}\hat{a}^{\dagger}_{\p_b}\hat{a}_{\p_a}\hat{a}_{\p_b}\rangle&= &\sum_{N=0}^{\infty}P_N  \int  \int_{\phi=0}^{2\pi} D_N[J]\, 
\sum_{n,n',m,m'=1}^{N} e^{i(\phi_n+\phi_m-\phi_{n'}-\phi_{m'})}e^{ip_a\cdot(x_n-x_{n'})}e^{ip_b\cdot(x_m-x_{m'})}\nonumber\\
&\times& \tilde{J}_0^*(p_a-p_{n'})\tilde{J}_0^*(p_b-p_{m'})\tilde{J}_0(p_b-p_{m})\tilde{J}_0(p_a-p_{n}).
\eeq
The phase averaging leaves three non-vanishing contributions: $n=n', m=m'$; $n=m', m=n'$ and finally the contribution which corresponds to emission of particles from the same source $J_0$, namely $n=n'=m=m'$. The first two contributions factorize {\it \`a la} Wick into products of the single particle averages similar to Eq. (\ref{adaga}). The third contribution cannot be written in this form, and we will call it ``irreducible''. It is affected by the structure of the current $J_0$. In fact, the coherent state approximation to an individual emitter Eq. (\ref{coherent}) is not always appropriate. Corrections to this classical structure of the source do not affect the factorizable contributions in Eq. (\ref{4a}), as those contributions depend only on the single particle spectrum of an individual emitter. They do however affect the 
irreducible contribution, since it probes possible correlations within a single coherent emitter $J_0$. However when the average number of coherent emitters $ N_S$ is large, the irreducible contribution is suppressed by the factor $1/N_S$. We will come back to this point later in Sec. \ref{graphs}.

All in all, the two-particle inclusive spectrum can be written as 
\beq
\langle \hat{a}^{\dagger}_{\p_a}\hat{a}^{\dagger}_{\p_b}\hat{a}_{\p_a}\hat{a}_{\p_b}\rangle&= &\sum_{N=0}^{\infty}P_N\int \left\{\prod_{i=1}^N d^4x_i\, d^4p_i \, 
n(x_i,p_i)\right\}\,\sum_{n\neq m}^N\bigg\{ |\tilde{J}_0(p_a-p_n)|^2 |\tilde{J}_0(p_b-p_m)|^2\nonumber\\
&&+ e^{i(p_a-p_b)\cdot(x_n-x_m)}\tilde{J}_0^*(p_a-p_m)\tilde{J}_0^*(p_b-p_n)\tilde{J}_0(p_b-p_m)\tilde{J}_0(p_a-p_n)\bigg\}\,+\,M^{irreducible}\,,
\eeq
where the $M^{irreducible}$ is defined as the $n=n'=m=m'$ contribution in Eq. (\ref{4a}). Thus, the two-particle spectrum reads
\beq\label{p2}
\P_2(\p_a,\p_b)= \frac{\langle N(N-1)\rangle }{\langle N \rangle^2}\bigg\{ \langle |\tilde{J}(\p_a)|^2\rangle \langle |\tilde{J}(\p_b)|^2\rangle + |\langle \tilde{J}^*(\p_a)\tilde{J}(\p_b)\rangle|^2\bigg\}+\frac{1}{N_S}\P_2^{irreducible},
\eeq
where we have indicated explicitly the $1/N_S$ suppression factor in  front of the irreducible contribution.
Neglecting for the moment the irreducible contribution, the single and two-particle distributions can be written in terms of the so called "emission function" $S(x, K)$  defined as
\beq
S(x,K)= \int \frac{d^4y}{2(2\pi)^3}e^{-iK\cdot y} \left \langle J^*\left(x+\frac{1}{2}y\right) J\left(x-\frac{1}{2}y\right) \right \rangle,
\eeq
and closely related to the Wigner distribution of the produced particles \cite{wigner}.
The correlators of the source functions that appear in both single and two-particle distributions are written in momentum space. Fourier transforming them, we write
\beq
\tilde{J}^*(\p_a) \tilde{J}(\p_b)=\int \frac{d^4x_1}{(2\pi)^3}\frac{d^4x_2}{2\sqrt{E_aE_b}} e^{-ip_a\cdot x_1 +ip_b\cdot x_2} J^*(x_1)J(x_2).
\eeq
Setting $x=\frac{1}{2}(x_1+x_2)$, $y=x_1-x_2$, $q=p_a-p_b$ and $K=\frac{1}{2}(p_a+p_b)$, we get
\beq
\tilde{J}^*(\p_a) \tilde{J}(\p_b)=\int \frac{d^4x_1}{(2\pi)^3}\frac{d^4x_2}{2\sqrt{E_aE_b}} e^{-iq\cdot x -iK\cdot y}J^*\left( x+\frac{1}{2}y\right) J\left( x- \frac{1}{2}y\right).
\eeq
Hence, the two-particle correlation function can be written as 
\beq\label{cqk}
C(\q, \K) = 1+\frac{\left | \int d^4x\,  S(x,K)\,  e^{iq\cdot x}\right |^2}{\int d^4x S\left( x, K+\frac{1}{2}q\right) \int d^4x S\left( x, K-\frac{1}{2}q\right)}\ .
\eeq
The second term in this expression expresses the HBT correlations, which arise due to the large number of incoherent sources that constitute the emitter.

The qualitative features of this expression are easy to understand. As a simple example let us take the individual sources to be static and distributed inside some radius $R$. For simplicity we take the actual distribution to be Gaussian, $n(x_i)\propto \exp\{-\frac{x_i^2}{2R^2}\}$. Also for simplicity we will assume that each source has a Gaussian profile with a radius $a$ much smaller than $R$, $J_0(x)\propto  \exp\{-\frac{x^2}{2a^2}\}$.
For the (square root of) the numerator of Eq. (\ref{cqk}) we obtain
\beq
&&\int d^3x_ie^{-\frac{x_i^2}{2R^2}}\int d^3xd^3 y e^{-iqx-iKy}e^{-\frac{1}{2a^2}\left[(x-x_i-\frac{y}{2})^2+(x-x_i+\frac{y}{2})^2\right]}\nonumber \\
&=&
\int d^3xd^3 y  d^3x_ie^{-iqx-iKy}e^{-\frac{x_i^2}{2R^2}}e^{-\frac{(x_i-x)^2}{2a^2}}e^{-\frac{y^2}{4a^2}}.
\eeq
As long as $a\ll R$, the second Gaussian acts as a delta function, not letting $x_i$ to stray too far from $x$, and thus the integral over $x_i$ gives
\beq
\int d^3xd^3 y e^{-iqx-iKy}e^{-\frac{x^2}{2R^2}}e^{-\frac{y^2}{4a^2}}\propto e^{-\frac{q^2R^2}{2}}e^{-K^2a^2}.
\eeq
In this simple model the HBT signal (the correlation between particles at different momenta) has a width $\sim 1/R$ in the momentum difference which does not depend on the total momentum of the pair $K$:
\beq\label{cqk1}
C(\q, \K) = 1+e^{-{\bf q}^2R^2}\,.
\eeq

In reality the shape of the emitter is more complicated and has to be modelled in some way. Also parts of the freeze-out surface are moving, and the source only exists for a finite time. All this introduces dependences on $K$ as well as on the distribution of $p_i$ as well. However, for our purposes the simple example above suffices since it illustrates the basic physics of the HBT correlations.

So far this has been a review of the standard theory of HBT. Our goal, though, is to understand what kind of HBT correlations one expects in the CGC picture of the collision. There are several important aspects which set this picture apart from the one that we have just described.

 First, one is interested in the emission of gluons and correlations between emitted gluons. For the sake of the argument we are going to assume local parton-hadron duality and forget about hadronization corrections. 
 
 Second, in the CGC picture the collision is boost-invariant. The colliding objects are strongly Lorentz contracted, and they overlap with each other only for an infinitesimally short time. The emitter exists only for this very short time, and this has a profound effect on the nature of the HBT correlations. 
 
Additionally, since we are discussing p-A scattering, we assume that the saturation momentum in the nucleus is much larger than that of the proton. Recall that the inverse saturation momentum is the transverse size of the ``patch'' of a hadron over which the color is correlated. Thus, different emitter points separated by distances greater than $Q_s^{-1}$ are uncorrelated in color. This random color orientation of different patches plays the role of the random phase of the individual sources in the previous discussion.

In the next section we discuss how these features are reflected in the HBT correlations between emitted gluons.

\section{The Gluon HBT}
\label{cgc}

As declared above, our goal is to understand gluon correlations. In this section we discuss the HBT signal for gluon emission in close analogy to the discussion of the previous section, highlighting its unique features. In the next section we will show that this discussion is closely paralleled by the glasma graph
calculation, which will allow us to identify the HBT contribution, with all its distinguishing properties, in the current calculations.

 \subsection{The gluon correlation function}
 
Like pions, gluons are emitted from the interaction region. However, gluon fields are real  and consequently the emitter current is real in coordinate space
(HBT of neutral pions would have to be treated similarly). 
Additionally, gluons carry an adjoint color index.
Thus, in the same spirit as in the previous section, let us consider a superposition of $N$ classical sources independently emitting gluons. Each source is translated to a different position $x_i$. The lack of coherence between the sources is encoded in a set of random adjoint unitary matrices $(U_A^i)_{ab}$, which determine the overall $SU(N_c)$ ``phase'' of each source. 

The time and longitudinal coordinate dependence of all sources is identical, and resides in the factor $\delta(x^+)\delta(x^-)$. Therefore, hereafter the plus and minus light-cone components of all four-coordinates must bet set to 0. 
The exact delta function nature of the source is of course an approximation. In actual fact, even at very high energy the gluon distribution in a hadronic wave function has a rapidity dependent structure. This is reflected in the Balitski-Fadin-Kuraev-Lipatov \cite{bfkl} (or Jalilian-Marian-Iancu-McLerran-Weigert-Leonidov-Kovner \cite{jimwlk}) evolution of the gluon distributions with rapidity. However, the rate of the rapidity evolution is proportional to the strong coupling constant $\alpha_s$. Consequently, as long as the rapidity interval between the observed particles in the final state is smaller than $1/\alpha_s$, approximate boost invariance holds and the source can be approximated by the delta function in the longitudinal direction. In the rest of this paper we will use this approximation.

The emitter source function is
\beq\label{jg}
\tilde{J}_a(\p)=\sum_{i=1}^N \; \left(U^i_A\right)_{ab} \; e^{ip\cdot x_i} \; \tilde{J}_{0\,b}(\p) \; ,
\eeq
with
\beq \tilde{J}_{0\,b}^{* }(\p)=\tilde{J}_{0\,b}(-\p).
\eeq
In this section, for simplicity of notation, we disregard the fact that the gluon field, and therefore also the gluon source, carry a transverse Lorentz (polarisation) index. This index will be restored in the next section.

The momentum $\p$ in the expression Eq. (\ref{jg}) and thereafter is the two-component transverse momentum. The independence of the source function on the longitudinal momentum is an immediate consequence (upon Fourier transform) of the localisation of the source in time and longitudinal coordinate. For the same reason we have not allowed for any transverse motion of parts of the emitter.

As before, we take (\ref{dm}) for the density matrix operator at a fixed value of the source $J^a$ with the coherent state expression
\beq\label{coherentg}
| J \rangle= \exp\left(i \int d^2p\, \tilde{J}_a(\p)\,  \left[\hat{a}^{\dagger}_{ a}(\p)+ \hat{a}_{a}(-\p)\right]\right) | 0\rangle.
\eeq 
Here, the state $|0\rangle$ is the vacuum of the gluon Fock space.
Since the source does not depend on the longitudinal momentum, the gluon creation and annihilation operators $\hat a^\dagger$, $\hat a$ here are integrated over rapidity \cite{KL1}. In terms of the fundamental rapidity dependent gluon creation and annihilation operators $\hat{a}^{(\dagger)}_{a}(\eta,\p)$, one has $\hat{a}^{(\dagger)}_{a}(\p)\equiv \int d\eta \, \hat{a}^{(\dagger)}_{a}(\eta,\p)$. This integration over rapidity is technically equivalent to approximating the source by a delta function in the longitudinal coordinate, and is valid with the same degree of accuracy.

The immediate consequence of this is that both the single gluon spectrum and the inclusive two gluon spectrum do not depend on the rapidity of the emitted gluons. This is precisely the form of the density operator used in the glasma graph calculations of gluon correlations.

The corresponding ensemble average of an arbitrary operator $\hat{O}$ is given as 
\beq
\tr( \hat{\rho} \, \hat{O} )=\sum_{N=0}^{\infty} P_N \int D_N[J]\, \langle J [N;\{x,U_A\}] | \hat{O} |  J [N;\{x,U_A\}] \rangle.
\eeq 
For the case at hand, the measure is
\beq
D_N[J]=\prod_{i=1}^N d^2x_i  \,n(x_i)\, [dU_A^i]
\eeq
with $\left[ dU\right]$ being the Haar measure,  $\int \left[ dU\right] =1$. 

One can calculate the single gluon correlator in a straightforward manner:
\beq
\langle \hat{a}^{\dagger}_{a} (\p)\hat{a}_b(\k)\rangle=\sum_{N=0}^{\infty}P_N\int D_N[J]\, \sum_{n,n'=1}^{N} ( U_A^n)_{bd} ( U^{\dagger\,n'}_{A})_{ca} \, e^{ik\cdot x_n-ip\cdot x_{n'}}\tilde{J}_{0\, c}^{*}(\p)\tilde{J}_{0\, d}(\k).
\label{single_particle_g}
\eeq
The integration over the unitary matrices can be performed by using the following formulae in the adjoint representation:
\beq
\int \left[ dU\right](U)_{ab}=0,\ \ \int \left[ dU\right](U)_{bd}(U^{\dagger})_{ca}= \frac{\delta_{ba}\delta_{dc}}{N_c^2-1}\ .
\eeq
Due to the orthogonality of the unitary matrices only the diagonal terms in the double sum over $n$ and $n'$ contribute. Thus, Eq.\eqref{single_particle_g} reads
\beq
\langle \hat{a}^{\dagger}_{a} (\p)\hat{a}_b(\k)\rangle&=&\sum_{N=0}^{\infty}P_N\int D_N[J]\, \sum_{n=1}^{N} ( U_A^n)_{bd} ( U^{\dagger\,n}_{A})_{ca} \, e^{i(k-p)\cdot x_{n}}\tilde{J}_{0\, c}^{*}(\p)\tilde{J}_{0\, d}(\k)\nonumber\\
&=& \, \frac{\delta_{ab}}{N_c^2-1}\langle \tilde{J}^{*}(\p)\tilde{J}(\k)\rangle    
\eeq
where, as before, we have defined for convenience
\beq\label{jj}
\langle \tilde{J}^{*}(\p)\tilde{J}(\k)\rangle \equiv N_S\int d^2x \,n(x)e^{-i(p-k)\cdot x}\,\tilde{J}_{0\, a}^{*}(\p)\tilde{J}_{0\, a}(\k), \ \ 
\langle \tilde{J}(\p)\tilde{J}(\k)\rangle \equiv N_S\int d^2x \,n(x)e^{-i(p+k)\cdot x}\,\tilde{J}_{0\, a}(\p)\tilde{J}_{0\, a}(\k).
\eeq
 For the single gluon spectrum, this gives
\beq
\langle \hat{a}^{\dagger}_{a} (\p)\hat{a}_a(\p)\rangle= \, \langle\tilde{J}^{*}_c(\p)\tilde{J}_c(\p)\rangle.
\eeq
An interesting and distinct property of the gluonic density operator, is that it yields a non-vanishing correlator of two-gluon annihilation operators, as well. This is a consequence of the reality of the gluon field (and source) in coordinate space. A simple calculation yields
\beq
\langle \hat{a}_{a} (\p)\hat{a}_b(\k)\rangle&=&-  \, \frac{\delta_{ab}}{N_c^2-1} \langle \tilde{J}(\p)\tilde{J}(\k)\rangle,  \\
\langle \hat{a}^{\dagger}_{a} (\p)\hat{a}^{\dagger}_{b}(\k)\rangle&=&- \, \frac{\delta_{ab}}{N_c^2-1} \langle \tilde{J}^*(\p)\tilde{J}^*(\k)\rangle  .
\eeq
The non-vanishing of these correlators is the consequence of the fact that the gluon operator does not carry any abelian conserved charge. Thus, the gluon density matrix, which in this case is a simple coherent state, does not possess a sharp "gluon number" but rather contains a superposition of Fock states with different numbers of gluons.

%
%
%
For the two-particle spectrum we have
\beq
\langle \hat{a}^{\dagger}_{a}(\p) \hat{a}^{\dagger}_{b}(\k) \hat{a}_a(\p)\hat{a}_b(\k)\rangle&=& \sum_{N=0}^{\infty}P_N\int D_N[J]\,\sum_{n,n',m,m'=1}^N(U_A^ n)_{ac}(U_A^m)_{bd}(U^{\dagger\,n'}_A)_{c'a}(U^{\dagger\,m'}_A)_{d'b}\nonumber\\
&\times& e^{ip\cdot (x_n-x_{n'})} e^{ik\cdot (x_m-x_{m'})} \tilde{J}_{0\, c'}^{*}(\p) \tilde{J}_{0\, c}(\p)\tilde{J}_{0\, d'}^{*}(\k)\tilde{J}_{0\, d}(\k).
\eeq
In the limit of very large number of incoherent sources the averaging over the $SU(N_c)$ phases picks out three non-vanishing contributions, corresponding to $(n=n';\ m=m')$, $(n=m';\ m=n')$ and $(n=m;\ n'=m')$. The first two are similar to those discussed in the previous section, while the last one is new and corresponds to Wick contracting two creation operators and two annihilation operators:
\beq
\langle \hat{a}^{\dagger}_{a}(\p) \hat{a}^{\dagger}_{b}(\k) \hat{a}_a(\p)\hat{a}_b(\k)\rangle&=& \sum_{N=0}^{\infty} P_N \int  \left\{\prod_{i=1}^{N} d^2x_i \ n(x_i) \right\}\,
\frac{1}{(N_c^2-1)^2}
\sum_{n\neq m}^{N}\Bigg\{
(N_c^2-1)^2\, \left | \tilde{J}^{*}_{0\, a}(\p) \right|^2 \left | \tilde{J}^{*}_{0\, b}\k) \right|^2\nonumber\\
&+& (N_c^2-1)\, e^{i(p-k)\cdot x_n }\, e^{-i(p-k)\cdot x_m}\, \left| \tilde{J}^{*}_{0\, a}(\p)\tilde{J}_{0\, a}(\k) \right | \, \left| \tilde{J}^{*}_{0\, b}(\k)\tilde{J}_{0\, b}(\p) \right |\, \nonumber\\
&+& (N_c^2-1)\, e^{i(p+k)\cdot x_n }\, e^{-i(p+k)\cdot x_m}\, \left| \tilde{J}^{*}_{0\, a}(\p)\tilde{J}^{*}_{0\, a}(\k) \right | \, \left| \tilde{J}_{0\, b}(\k)\tilde{J}_{0\, b}(\p) \right | \Bigg\}.
\eeq
Thus, the two-particle spectrum reads
\beq
\P_2(\p,\k)=\frac{\langle N(N-1)\rangle }{\langle N\rangle ^2} &&\Bigg\{ \left\langle \left| \tilde{J}(\p) \right |^2 \right \rangle  
 \left\langle \left| \tilde{J}(\k) \right |^2 \right \rangle \nonumber\\
 && +\frac{1}{N_c^2-1}\Bigg[ \left | \left\langle \tilde{J}^{*}(\p)\tilde{J}(\k)\right\rangle \right |^2 
 + \left\langle \tilde{J}^{*}(\p) \tilde{J}^{*}(\k)\right\rangle  \left\langle \tilde{J}(\p) \tilde{J}(\k)\right\rangle \Bigg] \Bigg\}.
\eeq
If the number of independent emitters is finite, an additional irreducible term is present in the spectrum:
\beq\label{p22}
\P_2(\p,\k)&=&\frac{\langle N(N-1)\rangle }{\langle N\rangle ^2} \Bigg\{ \left\langle \left| \tilde{J}(\p) \right |^2 \right \rangle  
 \left\langle \left| \tilde{J}(\k) \right |^2 \right \rangle +\frac{1}{N_c^2-1}\Bigg[ \left | \left\langle \tilde{J}^{*}(\p)\tilde{J}(\k)\right\rangle \right |^2 
 + \left\langle \tilde{J}^{*}(\p) \tilde{J}^{*}(\k)\right\rangle  \left\langle \tilde{J}(\p) \tilde{J}(\k)\right\rangle \Bigg] \Bigg\}\nonumber\\
 &+& \frac{1}{N_S}\P_2^{irreducible}.
\eeq
Defining the emission function 
\beq
S(x,K)= \int \frac{d^2y}{2(2\pi)^3}e^{-iK\cdot y} \left \langle J_a\left(x+\frac{1}{2}y\right) J_a\left(x-\frac{1}{2}y\right) \right \rangle,
\eeq
we can write the normalized correlation function as
 \beq\label{cg}
C(\q,\K)= 1+ \frac{1}{N_c^2-1}
\frac{\left | \int d^2x\,  S(x,K)\,  e^{iq\cdot x}\right |^2+\int d^2x S\left( x, \frac{q}{2}\right)e^{-2iK\cdot x} \, \int d^2x S\left( x, \frac{-q}{2}\right)e^{2iK\cdot x}}{\int d^2x S\left( x, K+\frac{1}{2}q\right) \int d^2x S\left( x, K-\frac{1}{2}q\right)}\ ,
\eeq
where we have neglected the irreducible term. As stated above,  in all equations in this Subsection the plus and minus light-cone components of all coordinate vectors must bet set to zero.

Eq. (\ref{cg}) differs from Eq. (\ref{cqk}) in two aspects. First, the number of gluons is $N_c^2-1$, and this is the origin of the suppression factor in front of the correlated term in Eq. (\ref{cg}). Second, gluon fields are real while the discussion in the previous section was for charged fields. As a result, Eq. (\ref{cg}) contains an extra correlation term - the second term in the numerator. As we will show below, this term contributes to the HBT enhancement  between gluons with opposite transverse momenta and in this sense has a distinct signature.
\subsection{The transverse structure of the source and the peculiarities of the gluon HBT}

The transverse structure of the source $J_a$ is what determines the HBT signal in Eq. (\ref{cg}). To understand it better, recall that the focus of our discussion is p-A collisions. The projectile proton carries into the collision region its distribution of classical gluon fields $b_a^i(x)$. This distribution in the proton wave function has some spatial size $R$, which we identify with the gluonic size of the proton. Thus a typical gluon field configuration in the proton 
is a slowly varying function of the transverse coordinate, which varies on the scale $R$. 

The effect of the collision with the target nucleus is to rotate the field $b_a^i(x)$ ($i$ is the transverse index) by an $x$-dependent $SU(N_c)$ adjoint matrix $U(x)$, so that immediately after the interaction the gluon field is 
$U_{ab}(x)b_b^i(x)$. As we will see in the next section, in the CGC approach the role of the source $J$ for the soft gluon emission is played to a good approximation by the gluon field at the instant following the interaction. Thus the source function that enters the calculation is closely related to the classical field,
\beq\label{sou}
J_a^i(x)\approx U_{ab}(x)\,b_b^i(x).
\eeq
The averaging over the event ensemble in the calculation of the correlator $\langle JJ\rangle$ amounts to averaging over the distribution of the
proton fields $b(x)$ and also over the distribution of the unitary matrices $U(x)$. The latter distribution is determined by the target wave function. Analogously to Eq. (\ref{jj}), we write
 \beq
 \langle J(x){J}(y)\rangle \equiv  \langle\langle J(x){J}(y)\rangle_{b(x)}\rangle_{U(x)}\,. 
\eeq

Assuming that the nucleus has a large saturation scale $Q_s\gg R^{-1}$ and that the eikonal scattering matrix $U(x)$ typically varies on the spatial scale of $Q_s^{-1}$, we conclude that  the source $J_a$ is color correlated only on distance scales of order $Q_s^{-1}$. The salient features of spatial structure of such source can be understood as visualizing the typical configuration $J_a(x)$ as a collection of independent color sources, each having a fixed orientation in color space and the transverse size $Q_s^{-2}$. The relative color orientation between the individual independent color sources in the event ensemble is completely random. The total area taken up by the source $J_a(x)$ is $R^2$, thus the number of independent sources is $N_S\sim Q_s^2R^2$. 

Schematically (we will be more precise in the next section) we can encode the previous discussion in the following form of the source:
\beq\label{sou}
J_a(x)=\sum_{i=1}^{N}U_{ab}^i n_b J_0(x-x_i),
\eeq
where $n_a$ is a fixed, unit length color vector common to all the individual sources, and the sources $J_0(x)$ have a range of $Q_s^{-1}\ll R$. The exact spatial dependence of the individual source $J_0(x)$ is not important for our purposes. It affects the "irreducible" contribution to the correlation function, but has no effect on the HBT signal.
As a simple model for the source  distribution we can again use a Gaussian like in the previous section:
\beq\label{gauss1}
n(x_i)\propto e^{-\frac{x_i^2}{2R^2}}.
\eeq
Note however that, as opposed to the previous section, the Gaussian profile in Eq. (\ref{gauss1}) involves only transverse coordinates.
 For the correlation function we then obtain
\beq
C(\q,\K)=1+\frac{1}{N_c^2-1}\left[ e^{-R^2\q^2}+e^{-R^2\K^2}\right].
\eeq
The first term within the square bracket in this expression is the HBT signal of the``usual'' kind. It leads to maximum of the correlation when the momenta of the two emitted gluons are almost equal to each other (with accuracy of $R^{-1}$). The second term is somewhat unusual, since it maximises the correlation when the momenta of two emitted gluons are antiparallel to each other (with the same accuracy). This property of the gluon correlation function, namely that it is symmetric under the reversal of the direction of momentum of one of the gluons, has been discussed extensively in the literature, and is the consequence of the reality of the gluon production amplitude. This unusual term also constitutes part of the {\it bona fide} HBT correlation.

To summarise this section we list the interesting properties of the gluon HBT correlation:
\begin{itemize}
\item The correlation is long range in rapidity - it is equally strong when the rapidities of the two gluons are equal or when the difference between the two rapidities is large.
\item The correlation is symmetric under reversal of the direction of the transverse momentum of one of the gluons. Thus it is strongest when the transverse momenta of the two gluons are either parallel or antiparallel.
\item The HBT radius is of the order of the inverse gluonic size of the proton $R^{-1}$.
\item The HBT signal dominates the correlation function (at small momentum difference) when the number of incoherent emitters is large, $N_S=Q_s^2R^2\gg 1$.

\end{itemize}

\section{The HBT in glasma graphs}
\label{graphs}

In this section we identify the diagrams responsible for the HBT correlations in the ambient glasma graph calculation. 
Recall that in the eikonal approach the important contribution to the two gluon inclusive cross section in p-A collisions is given by
\beq\label{ja}
\langle out\vert a^{\dagger \, i}_a(p)a^{\dagger\, j}_b(k)a^i_a(p)a^j_b(k)\vert out\rangle
 &=&\int dU \,W^T[U]\int D\rho \,W[\rho]\ \langle 0\vert  e^{i\int _x
 J_a^i(x)\left[a^{\dagger \, i}_a(x)+a_a^i(-x)\right]}
a^{\dagger \, i}_a(p)a^{\dagger \, j}_b(q)a^i_a(p)a^j_b(q)\nonumber \\
&\times&
 e^{-i\int _x
 J_a^i(x)\left[a^{\dagger\, i}_a(x)+a_a^i(-x)\right]}\vert 0\rangle 
 \eeq
 (all coordinates and momenta in this Section are transverse and the corresponding integrals two-dimensional). Here, the state $|0\rangle$ is the vacuum of the soft-gluon Fock space in the presence of the  "valence gluon" - a color charge density  $\rho_a(x)$, the Weizs\"acker-Williams field of the incoming projectile is given by $b_a^i(x)=\int d^2y\frac{\partial^i}{\partial^2}(x,y)\rho_a(y)$, and the color charge density configurations are distributed according to the probability density $W[\rho]$, while those of the color matrices in the target according to $W^T[U]$. The current $J$ is given by 
 \beq\label{sourceg}
 J_a^i(x)=U_{ab}(x)b^i_b(x)-\int d^2y\frac{\partial^i}{\partial^2}(x,y)U_{ab}(y)\rho_b(y).
 \eeq
 Note that Eq. (\ref{ja}) is not the complete expression for inclusive two gluon production, but only the part where the two gluons are produced from two different Pomerons. This is the part of the production cross section which leads to the correlated result, and the only one that we will consider in this paper.

The first observation is that the expression in Eq. (\ref{ja}) is exactly of the type discussed in the previous section. The averaging of the gluonic creation and annihilation operators is performed over a classical coherent state specified by the source Eq. (\ref{sourceg}), with subsequent averaging over the ensemble of sources distributed with some probability density $W$.
The source is slightly different from the one we discussed in the previous section Eq. (\ref{sou}), but not significantly so. The first term in Eq. (\ref{sourceg}) is precisely the source of Eq. (\ref{sou}). The second term is additional, however it is well known \cite{KM} that the emission of gluons with large transverse momentum (of the order of $Q_s$) is dominated by the first term in Eq. (\ref{sourceg}). Thus, the presence of the additional term in Eq. (\ref{sourceg}) does not change significantly our expectations based on the discussion of the previous section.

To calculate the two gluon inclusive cross section one has to average over the distribution of the color charge density, and also over the distribution of target fields $U_{ab}(x)$. The latter averaging introduces $Q^{-1}_s$ as the typical transverse scale of variation of $U(x)$. In current implementation one uses a Gaussian weight (the so called McLerran-Venugopalan model \cite{mv}) for averaging over the color charge density, and a similar factorizable distribution for the target average.

\begin{figure*}[htp]
\includegraphics[width=\textwidth]{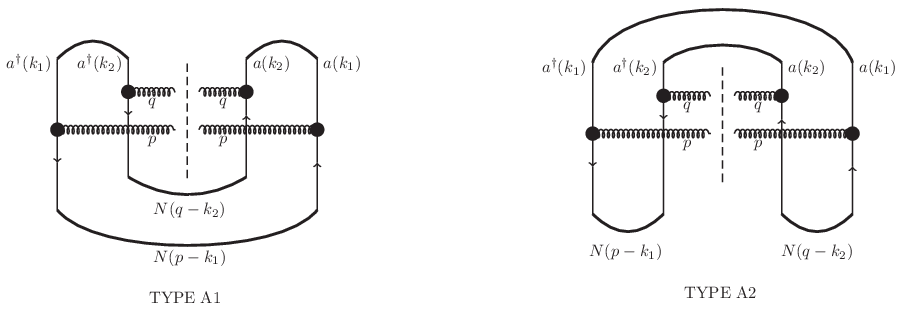}
\includegraphics[width=15cm]{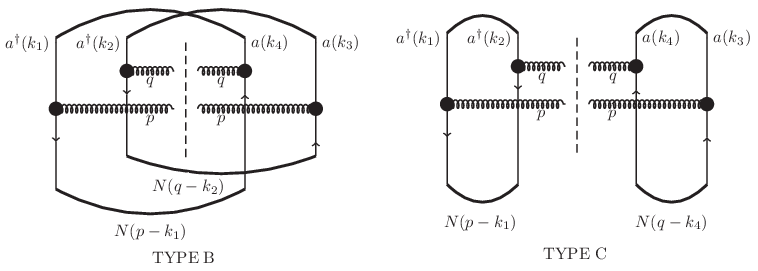}
\caption{\label{4g}Glasma graphs for the irreducible contribution (Types A1 and A2) and for HBT correlations (Types B and C).  $N(p-k)$ is proportional to the expectation value of the product of two $U$ matrices, which has the meaning of the probability that the incoming gluon with transverse momentum $k$ acquires transverse momentum $p$ after scattering. }
\end{figure*}

After taking the expectation value over the soft gluon Hilbert space, the expression in Eq. (\ref{ja}) can be written as
\beq
\langle out\vert a^{\dagger\, i}_a(p)a^{\dagger \, j}_b(k)a^i_a(p)a^j_b(k)\vert out\rangle=\langle J^{\dagger \, i}_a(p)J^{\dagger \, j}_b(k)J^i_a(p)J^j_b(k)\rangle_{\{\rho,U\}}\ ,
\eeq
where the averaging over the distributions of $\rho$ and $U$ has to be performed. Using the factorizable model for averaging, this can be represented in terms of simple diagrams \cite{ddgjlv}, which were used in \cite{DV} to calculate the gluon correlations numerically. 
The diagrams are depicted in Fig. \ref{4g}. 

It is easy to understand what is the role of the different diagrams in the framework that we have discussed in the last section (see also the discussions in \cite{BE}). First off, the diagrams that contain contributions of Eq. (\ref{cg}) are the ones shown in Fig. \ref{4g} named B and C. If in the calculation of the diagrams one assumes translational invariance of the projectile wave function, the loop integrals 
lead to momentum $\delta$ functions:
\beq \label{trans}
B\propto \delta^{(2)}(q-p),\ \   C\propto \delta^{(2)}(q+p).
\eeq
Relaxing the translational invariance approximation, the delta functions are smeared over a scale of the size  $R$ of the projectile\footnote{Note that  the gluon fields in the proton can be correlated on a distance scale $r<R$. This corresponds to a "domain" picture where the proton contains several domains of color fields \cite{correlations}. Nevertheless, as long as $r\gg Q_s^{-1}$ the scale that determines the HBT radius is the overall gluonic proton size $R$. Physically this is obvious since the only determining factor for HBT is the structure of the emitter, Eq. (\ref{sou}). In turn, the size of the correlated regions in the emitter is determined by the smallest spatial inhomogeneity scale which is $Q_s^{-1}$ .  Mathematically, within the "glasma  graph" calculation, the reason is that the momentum $q-p$ in Eq. (\ref{trans})  is conjugate to the center of mass coordinate in the gluon correlator $\langle a^\dagger(x)a(y)\rangle $. The domain picture corresponds to the correlator of the type $\langle a^\dagger(x)a(y)\rangle \sim f_r(x-y)g_R((x+y)/2)$, where the function $f_r$ vanishes when $|x-y|>r$, while $g_R$ vanishes when $|x+y|/2>R$. The smearing of the $\delta$-function in the first relation in Eq. (\ref{trans}) clearly happens on the scale $R$. The same conclusion is reached for the second relation in Eq. (\ref{trans}).}. Thus diagram B represents the second term in Eq. (\ref{cg}), while diagram C the third term in Eq. (\ref{cg}).
The identification of the diagrams B and C with the HBT correlation contribution is not new. It was known to the authors of \cite{ddgjlv}  (\cite{adrian}), and was also discussed subsequently in \cite{yury}. An earlier discussion of the diagrammatic representation of HBT can be found in \cite{capella}.
The peculiarities of the gluon HBT signal was however not discussed in these references.

It is also interesting to understand how the remaining diagrams, called Type A1 and A2, fit into the framework of our discussion. It was shown in \cite{BE} that the diagrams of Type A yield correlations due to Bose enhancement of gluons in the {\it incoming} wave function of the projectile and the target. This is the standard enhancement of the probability to find two incoming gluons in a hadronic wave function to have exactly the same transverse momentum. Due to momentum exchange during the scattering process, the enhancement is smeared over the momentum range of order $Q_s$, but the correlation still remains. As discussed in \cite{ddgjlv} and \cite{BE}, the contribution of these graphs to the correlation $C(\q,\K)$ is suppressed by the factor $1/(Q^2_sS)$, where $S$ is the area of the smaller hadron (the proton).  Identifying $S$ with $R^2$, we see that this suppression factor is precisely the factor $1/N_S$ of the previous section. Thus the Type A contribution is identified with the irreducible term ${\cal P}_2^{irreducible}$ in Eq. (\ref{p22}). Physically, this of course is clear, since the Type A contributions originate from correlations in the initial state, while the HBT correlations are due to independent uncorrelated emissions from the emitter formed immediately after the collision.

\section{Discussion}
\label{disc}

We have shown in this paper that the HBT signal in emission of gluons from the initial stages of p-A collisions has some interesting properties.

It is boost invariant, that is, the HBT radius in the longitudinal direction is very small. In the boost invariant approximation used here, this radius is strictly zero and the correlation extends to arbitrary rapidity difference between the two emitted particles. Naturally, one expects that, at very large rapidity differences where the quantum evolution is important, the HBT signal must drop down. This should only happen for relatively large rapidity difference between the two gluons $\Delta\eta\gtrsim 1/\alpha_s$.

In transverse space, the HBT radius is given by the gluonic radius of the proton. This, in principle, gives a possibility to directly measure the radius of the gluon distribution in the proton, assuming the signal is not washed out by final state effects. It is an interesting question what the gluonic radius of the proton is. The naive estimate would be of the order of the strong interaction radius $\sim 0.8-0.9$ fm. However there are many indications, particularly based on the DIS data, that the gluons in the proton are much more compact, taking up only about $0.3$ fm (see e.g. \cite{LL}). A direct measurement of this quantity in p-A collisions would be extremely interesting.

The HBT correlation is equally strong for two gluons with equal transverse momenta and momenta equal in magnitude but opposite in direction. This is a general feature of the gluon two-particle correlation.

\begin{figure*}[htp]
\includegraphics[width=10cm]{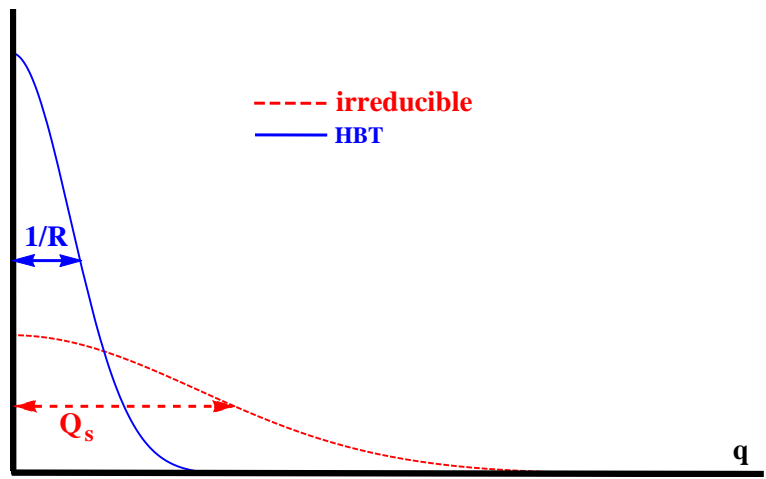}
\caption{\label{fig2}Schematic separation in $\q$ (or $\K$) of the uncorrelated HBT (solid line) and irreducible initial-state Bose (dashed line) signals. Horizontal and vertical scales are arbitrary.}
\end{figure*}

Finally, we have noted that the complete two gluon correlation within the glasma graph approach comprises two distinct contributions: the HBT and the irreducible initial state Bose enhancement. The relative importance of the two contributions is determined by the number of independent emitters $N_S=Q_s^2R^2$. The two parts of the signal have different nature and different properties. The irreducible part is suppressed by $1/N_S$, but leads to correlations whose width in the momentum space is determined by the saturation momentum $Q_s$. On the other hand the HBT signal is unsuppressed, but is much narrower in momentum space, with the width $R^{-1}$. The total signal that one expects is therefore a superposition of the two, and should have the general shape depicted on Fig. \ref{fig2}.  
It would be extremely interesting to see whether this double scale structure can be observed in the correlation. To do that, however, one would need to bin the signal into much narrower transverse momentum bin sizes than is currently done.

We note that the separation of the two gluon correlation function into the HBT and the irreducible part only makes sense if the gluonic size of the proton is much larger than the inverse saturation momentum of the nucleus. Otherwise $N_S\sim 1$ and the two signals become indistinguishable. Therefore one expects  that this characterization is better suited for high multiplicity events, which correspond to the largest values of $Q_s$.

Of course, the usual word of caution is due here. All of the above discussion disregarded possible effects of  final state interactions. Although the current thinking is predominantly that  final state effects are very important in p-A collisions, nevertheless, as noted in the beginning, the ridge description based on initial state correlations is quite successful in reproducing available data \cite{DV}. In this vein one may hope that the final state effects do not completely obliterate the initial state signal, including the HBT correlations, at least for high enough transverse momentum. Admittedly, since the HBT signal is expected to be quite narrow, it is also very fragile and will be affected the most by any transverse momentum smearing in the final state. This includes final state interactions, but also fragmentation effects \footnote{
In this connection we note that the HBT contribution from diagrams of Type B and C was included in the numerical calculations of \cite{DV}. In these works the naive momentum delta function (that arises in the translationally invariant limit) was smeared on the scale of the saturation momentum, which made the HBT signal indistinguishable from the irreducible Bose enhancement correlation. The motivation for such smearing is precisely the effects of possible momentum transfer by  final state interactions. }.
The fragility of the signal makes it rather tricky to predict in which kinematic regime it can be seen best. In particular, significantly increasing $Q_s$ also increases the number of produced particles and consequently the probability of final state interactions. On the other hand, for small $Q_s$ the difference between the HBT signal and the irreducible contribution to the correlation disappears. Thus, in order to see the signal it may be advantageous to study relatively forward production in p-A collisions in events with moderately high multiplicity.

\section*{Acknowledgments}

 We thank Adrian Dumitru  and Douglas Wertepny for useful discussions. We express our gratitude to the Department of Physics of Ben-Gurion University of the Negev, for warm hospitality during stays when parts of this work were done (TA, NA and AK) and for financial support as Distinguished Scientist Visitor (NA). 
 ML thanks the Physics Department of the University of Connecticut for hospitality. 
This research  was supported by the People Programme (Marie Curie Actions) of the European Union's Seventh Framework Programme FP7/2007-2013/ under REA
grant agreement \#318921; the DOE grant DE-FG02-13ER41989 (AK); the BSF grant \#2012124 (ML and AK); the Kreitman Foundation (GB);  the  ISRAELI SCIENCE FOUNDATION grant \#87277111 (GB and ML);  the European Research Council grant HotLHC ERC-2011-StG-279579, Ministerio de Ciencia e Innovaci\'on of Spain under project FPA2014-58293-C2-1-P, Xunta de Galicia (Conseller\'{\i}a de Educaci\'on and Conseller\'\i a de Innovaci\'on e Industria - Programa Incite),  the Spanish Consolider-Ingenio 2010 Programme CPAN and  FEDER (TA and NA).

\end{document}